\documentclass[a4paper,12pt]{article}

\usepackage[centertags]{amsmath}
\usepackage{amsfonts}
\usepackage{amssymb}
\usepackage{amsthm}
\usepackage{newlfont}
\usepackage[english]{babel}
\usepackage{enumerate}

\newcommand{\lam}{\lambda}

\newcommand{\nsp}{\overline{\mathbf{N}}}

\theoremstyle{plain}
  \newtheorem{theorem}{Theorem}[section]

\theoremstyle{definition}

\theoremstyle{remark}

\numberwithin{equation}{section}

 \DeclareMathOperator{\supp}{Supp}

\newcommand{\opunit}{\text{1}\kern-0.22em\text{l}}

\newcommand{\bsi}{{\boldsymbol i}}
\newcommand{\bsj}{{\boldsymbol j}}
\newcommand{\bsk}{{\boldsymbol k}}
\newcommand{\bsl}{{\boldsymbol l}}
\newcommand{\bsm}{{\boldsymbol m}}
\newcommand{\bsn}{{\boldsymbol n}}
\newcommand{\bso}{{\boldsymbol o}}
\newcommand{\bsp}{{\boldsymbol p}}

\newcommand{\bsL}{{\boldsymbol L}}

\newcommand{\beq}{ \begin{equation} }
\newcommand{\eeq}{ \end{equation} }
\newcommand{\bet}{ \begin{theorem} }
\newcommand{\eet}{ \end{theorem} }

\newcommand{\wpsi}{{\widehat{\Psi}}}

\author{
J.~P.~Cruz\thanks{Mathematics Departament, University of Aveiro, Portugal, e-mail: pedrocruz@ua.pt}
\and\
E.~L.~Lakshtanov\thanks{Mathematics Departament, University of Aveiro, Portugal, e-mail: lakshtanov@rambler.ru}
}

\title{EXPLICIT REPRESENTATION OF GREEN FUNCTION FOR 3D DIMENSIONAL EXTERIOR HELMHOLTZ EQUATION}
\date{}

\begin{document}
\selectlanguage{english}
\maketitle

\begin{abstract}
We have constructed a sequence of solutions of the Helmholtz equation
forming an orthogonal sequence on a given surface. Coefficients
of these functions depend on an explicit algebraic formulae from the
coefficient of the surface.  Moreover, for exterior Helmholtz
equation we have constructed an explicit normal derivative of the
Dirichlet Green function. In the same way the Dirichlet-to-Neumann
operator is constructed. We proved that normalized coefficients
are uniformly bounded from zero.
\end{abstract}


\tableofcontents

\section{Introduction}
Consider $\Omega \subset \mathbb R^3$ with Lipschitz boundary
$\partial \Omega$ and $k > 0$. The scattered field is given by
Helmholtz equation and radiation condition
\begin{equation}\label{helm}
\Delta \Psi(r)+k^2 \Psi(r)=0, \quad r \in \Omega'=\mathbb R^3
\backslash \Omega,
\end{equation}
\begin{equation}\label{Somm}
\int_{|r|=R} \left |\frac{\partial \Psi(r)}{\partial
|r|}-ik\Psi(r) \right |^2 dS = o(1), \quad R \rightarrow \infty,
\end{equation}
with Dirichlet boundary conditions,
\begin{equation}\label{dir}
\Psi(r) \equiv u_0(r), \quad r \in \partial \Omega, \quad u_0 \in
C(\partial \Omega).
\end{equation}
For example, in \cite{vantychonovsamarsky} is proved the existence
and uniqueness of the solution of (\ref{helm})-(\ref{dir}). A
function $\Psi(r)$ which satisfy mentioned conditions has
asymptotics
\begin{equation}\label{scamp}
\Psi(r)=\frac{e^{ik|r|}}{|r|} f(q)+o \left  (\frac{1}{|r|} \right
), \quad r \rightarrow \infty, \quad q=r/|r| \in S^2,
\end{equation}
where function $f(\theta,\varphi)=f(\theta,\varphi,k,u_0)$ is
called {\it scattering amplitude} and observable
$$
\sigma_T=\int_{S^2} |f(q)|^2 d\sigma(q)
$$
is called Total Cross Section, $\sigma$ is a square element of the
unit sphere.

A very important particular case of the boundary condition is
$$
u_0=e^{ik<r,\theta_0>}, \quad r \in \partial \Omega,
$$
which is the scattering of a plane wave with incident angle
$\theta_0 \in S^2$. The total momentum transmitted to the obstacle
is given by observable called Transport Cross Section (in a large
volume normalization)
$$
 \quad R=\int_{S^2} (1-<\! q,\theta_0 \!>)|f(q)|^2 d\sigma(q).
$$
%

Unfortunately, analytical expressions of these observables for
certain $k
> 0$ exist only for few
bodies of simple shapes (see
\cite{simsh,prolate}). Moreover, the scattering happens not
only by plane or spherical wave, but it could be caused, for
example, by arbitrary secondary radiation.

From another point of view there exists some numerical methods for
direct scattering calculation. One of them is based on a numerical
solution of integral equation (see \cite{crcol}). Another method,
developed by A.~Ramm and S.~Gutman in
\cite{AlRammPhysA}-\cite{AlRammSm}, allow to construct the Green
function and therefore to obtain solutions for arbitrary boundary
condition $u_0 \in C(\partial \Omega)$. The ground analytical
achievement by A.~Ramm and S.~Gutman is the so called Modified
Rayleigh Conjecture. In particularly, it follows that functions
$Y_{lm}(\theta,\varphi)h_l(k|r|)|_{\partial \Omega}$ (spherical
harmonics and spherical Hankel functions (see (\ref{hankdef}))
correspondingly) form a basis in the space $L_2(\partial
\Omega,dS)$.

The aim of current work is to express explicitly coefficients of
physical observables  and also for the normal derivative of the
Dirichlet-Green function. We found a constant uniform lower bound
for normalization coefficients (denominators) and we prove
convergence of all produced series.

Green’s function of the Laplacian in $\Omega'$ is given by
$$
\Delta_r G(r,t)+ k^2 G(r,t)=-\delta(r-t), \quad r,t \in \Omega',
\quad G \equiv 0, \quad r \in \partial \Omega,
$$
$$
\lim_{r \rightarrow \infty} \int_{|t|=R}\left |\frac{\partial
G(r,t)}{\partial n_t}-ikG(r,t) \right |^2 dS = o(1), \quad R
\rightarrow \infty.
$$
As it is known (evidence from Green formula)
$$
\Psi(r)=\int_{\partial \Omega}\frac{ \partial G(r,t)}{\partial
n_t} u_0(t)dS(t) , \quad r \in \overline{\Omega'}
$$
In this paper we shall express kernel $\frac{ \partial
G(r,t)}{\partial n_t}$ explicitly through coefficients of the
surface $\partial \Omega$ and also as measures $C$, $\sigma_{T}$,
$R$ and $\{A_{lm}\}$, where $A_m, 0\leq |m| \leq l$ are
coefficients of representation
\begin{equation}\label{defAm}
f(\theta,\varphi)=\sum_{l=0}^\infty \sum_{|m|\leq l} A_{ml}
Y_{lm}(\theta,\varphi),
\end{equation}
 where $Y_{ml}(\theta,\varphi)$ are  {\it spherical harmonics}.

Let $\nsp=\mathbf N \cup \{0\}$. Let $\mathcal L$ be the set of
indexes
$$
\mathcal L = \{ (l,m) : l,m \in \nsp, |m| \leq l \}
$$
and let $\bsl \in \mathcal L$ with $\bsl=(\bsl(1),\bsl(2))$. We also
we use the notation $\overline \bsl =(\bsl(1),-\bsl(2))$ and
set the  order: $(l,m)
> (p,r) \longleftrightarrow l>p \wedge  [(l=p) \vee (|m|>|r|)]
\wedge [(l=p) \vee (m>0) \vee (m=-r)]$, so
$\mathcal L =\{(0,0),
(1,0),(1,-1),(1,1),(2,0),(2,-1),(2,1),(2,-2),(2,2),(3,0),...\}$.
Let $\bso=(0,0)$ and let operations
$+$ and $-$ have the natural definition in $\mathcal L \times \mathbf
Z \rightarrow \mathcal L$ correspondingly to the introduced order.

\subsection{Surfaces with inverse radius-vector represented as finite combination of harmonics}

Let $\mathcal F$ be a subset of functions (multiindex) $\mathcal
L^{\nsp}$ which have finite support and let capacity be defined by
$$
 |d|=\sum_{\bsl} d(\bsl), \quad \supp{d} = \max\{\bsl : d(\bsl)
\neq 0 \}\,.
$$
Let $e_{\bsl} \in \mathcal F$ be defined as
$e_{\bsl}(\bsm)=\delta_{\bsl \bsm}$ (evaluated $1$ only when $\bsl=\bsm$).
Also set
$$
C^d=\frac{|d|!}{\prod_{\bsl} d(\bsl)!}, \quad a^d=\prod_{\bsl}
a_{\bsl}^{d(\bsl)},
$$
$$
I^d=\int_{0}^{\pi} \int_{0}^{2\pi} \prod_{\bsl} \left (Y_{\bsl}
(\theta,\varphi) \right )^{d(\bsl)} d \theta d\varphi.
$$

\begin{theorem}\label{th}Let a star shaped surface $\partial \Omega$
be given as a set $\{r=r(\theta,\varphi) \in \mathbb R^3, \theta
\in [0,\pi], \varphi \in [0,2\pi) \}$ where
\begin{equation}
|r(\theta,\varphi)|=\frac{1}{\sum_{\bsl \leq (N,N)} a_{\bsl}
Y_{\bsl}(\theta,\varphi)}
\end{equation}
with $N \geq 0$ and where $\{a_{lm}\}$ are coefficients. Then
\\ 1. Functions
$$
\wpsi_\bsn(r)=\sum_{\bsk \leq \bsn} c_{\bsn \bsk}
Y_\bsk(\theta,\varphi) h_{\bsk(1)}(k|r|), \quad \bsn \in \mathcal
L
$$
satisfy (\ref{helm}),(\ref{Somm}) and their restrictions $\{
\wpsi_\bsn |_{\partial \Omega} , \bsn \in \mathcal L\}$ form an
orthonormal basis in $L_2(\partial \Omega, d\theta d \varphi)$.
\\
Here
$$
c_{\bsn \bsn}=1/\lam_{\bsn}, \quad \bsn \in \nsp
$$
$$
\lam_\bso^2=g_{\bso \bso}>0,  \quad \lam_\bsn^2=g_{\bsn \bsn} -
\sum_{\bsk=\bso}^{\bsn-1} \left | \sum_{\bsp=\bso}^{\bsk}
\overline{c}_{\bsk \bsp} g_{\bsn \bsp} \right
 |^2 >0, \quad \bsn > \bso.
$$
We now define $g_{\bsi \bsj}$, $\widehat{h}_{nm}$, $c_{\bsn
\bsm}$. Let
$$
g_{\bsi \bsj}=(-1)^{\bsj(2)}\sum_{m=0}^{\bsi(1)+\bsj(1)}
\frac{1}{k^{m+2}}\left (\sum_{l=0}^{m} \widehat{h}_{\bsi(1)l}
\overline{\widehat{h}}_{\bsj(1)(m-l)} \right ) \sum_{d: |d| =m+2,
\supp{d} \leq
 (N,N)} C^d a^d I^{d+e_\bsi+e_{\overline \bsj }}
$$
where coefficients $\widehat{h}_{nj}$ are defined from the well
known representation for Hankel spherical functions \cite{morsef},
\begin{equation}\label{hankdef}
h_{n}(t)=\frac{e^{ikt}}{t}\frac{\sum_{j=0}^{n}\widehat{h}_{nj}t^{n-j}}{t^n}=
\frac{e^{ikt}}{t}\sum_{j=0}^{n}\frac{\widehat{h}_{nj}}{t^{j}},
\quad \widehat{h}_{n 0}=1, \quad t \neq 0,
\end{equation}
with
$$
 \widehat{h}_{nm}=\frac{i^m}{2^m} \prod_{p=1}^m (n+p) \cdot
\prod_{p=1}^m \frac {(n-m+p)}{p}, \quad 0<m\leq n;
$$
for $\bsm < \bsn$ we have
$$
c_{\bsn \bsm} = \frac{1}{\lam_\bsn} \left (
\sum_{\bsk=\bsm}^{\bsn-1}  \sum_{\bsp=\bso}^{\bsk}
\overline{c}_{\bsk \bsp} c_{\bsk \bsm} g_{\bsn \bsp} \right ).
$$
\\ 2. Consider an arbitrary function $u_0 \in L_2(\partial \Omega)$,
then
 we have
\begin{equation}\label{sigmat}
\sigma_T=\frac{1}{k^2} \sum_{\bsn=\bso}^\infty\sum_{\bsm \leq
\bsn} \left [\overline{c}_{\bsn \bsm} \left ( \sum_{\bsp \leq
\bsn} c_{\bsn \bsp} \overline{\widehat{u}}_{\bsp} \right )\left
(c_{\bsn \bsm} \sum_{\bsp \leq \bsn} \overline{c}_{\bsn \bsp}
{u}_{\bsp}+2\sum_{\bsm <\bsl < \bsn} c_{\bsl \bsm} \sum_{\bsp \leq
\bsl} \overline{c}_{\bsl \bsp} {u}_{\bsp}\right ) \right ],
\end{equation}
and coefficients of the scattering amplitude
\begin{equation}\label{amam}
A_\bsm=\frac{1}{k}\sum_{\bsn \geq \bsm} c_{\bsn \bsm} \sum_{\bsp
\leq \bsn} \overline{c}_{\bsn \bsp} {u}_{\bsp}, \quad \bsm \in
\mathcal L,
\end{equation}
where
$$
{u}_{\bsp}=\int_{0}^\pi \int_{0}^{2\pi} u_0(\theta,\varphi)
\overline{Y}_\bsp(\theta,\varphi) \overline{h}_{\bsp(1)}(k
|r(\theta,\varphi)|) d \theta d\varphi.
$$
\\
3. Moreover, exists numbers $C_i=C_i(k,\Omega), i=1,2$  such that
\begin{equation}\label{estth1}
c_{\bsn \bsk}\leq \frac{C_1}{\bsk(1)!}, \quad \bso \leq \bsk \leq
\bsn, \quad \bsk,\bsn \in \mathcal L.
\end{equation}
 Also
\begin{equation}\label{estth2}
 c_{\bsn
\bsm} \left | \sum_{\bsp=\bso}^\bsn \overline{c}_{\bsn \bsp}
u_\bsp \right | < \frac{C_1}{\bsm(1)!} |\widehat{u}_\bsn|, \quad
\widehat{u}_\bsn=(u_0,\wpsi_\bsn |_{\partial
\Omega})_{L_2(\partial \Omega, d\theta d\varphi)}=\sum_{\bsk \leq
\bsn} \overline{c}_{\bsn \bsk}u_\bsk \rightarrow 0,
\end{equation}
when $\bsn \rightarrow \infty$ and
\begin{equation}\label{estth3}
 \lam_\bsn > C_2, \quad \bsn \in \mathcal L.
\end{equation}
\\ 4.
We have  weak convergence of $\frac{\partial G}{\partial n_t}$:
$$
\frac{\partial G}{\partial n_t}(r,t)=\sum_\bsn \wpsi_\bsn(r)
\overline{\wpsi}_\bsn(t), \quad r,t \in \Omega' .
$$

\end{theorem}

{\bf Note 1.} Define $\{A_{\bsm}^\bsL, \bsm,\bsL \in \mathcal L\}$
where $A_\bsm^\bsL=0$ for $\bsL<\bsm$ and
$$
A_\bsm^\bsL=\frac{1}{k}\sum_{\bsL \geq \bsn \geq \bsm} c_{\bsn
\bsm} \sum_{\bsp \leq \bsn} \overline{c}_{\bsn \bsp}
\widehat{u}_p, \quad \bsL \geq \bsm.
$$
or, recursively
$$
A_\bsm^{\bsL}=A_\bsm^{\bsL-1}+\frac{1}{k}  c_{\bsL \bsm}
\sum_{\bsp \leq \bsL} \overline{c}_{\bsL \bsp} \widehat{u}_p.
$$
From Theorem \ref{th} and using the orthogonality of basis $\{
\wpsi_\bsn |_{\partial \Omega}\}$ we have
$$
|A_\bsm - A_\bsm^\bsL| \leq \frac{C}{\bsm(1)! } \left
\|u_0-\sum_{\bsp \leq \bsL} \widehat{u}_\bsp \right \|=o(1), \quad
\bsL \rightarrow \infty\,.
$$
Then
$$
\sigma_T^\bsL=\sum_{\bsn \leq \bsL} |A_{\bsn}^\bsL|^2,
$$
$$
| \sigma_T-\sigma_T^\bsL| \leq C \left \|u_0-\sum_{\bsp \leq \bsL}
\widehat{u}_\bsp \right \|=o(1), \quad \bsL \rightarrow \infty\,.
$$
For the case $u_0=e^{ikz}$, i.e. $\theta_0=(0,0,1)$, we have
\begin{equation}\label{ares}
R^\bsL=\sum_{\bsn \leq
\bsL}\sqrt{\frac{(\bsn(1)+\bsn(2))(\bsn(1)-\bsn(2))}{(2\bsn(1)+1)(2\bsn(1)-1)}}Re(A_{\bsn}
\overline{A}_{\bsn+e_{(1,0)}})
\end{equation}
$$
|R-R_\bsl| \leq  \|u_0-\sum_{\bsp \leq \bsL} \widehat{u}_\bsp
\|=o(1), \quad \bsL \rightarrow \infty .
$$
Formula (\ref{ares}) is a simple corollary of equality
$xP_n(x)=\frac{n+1}{2n+1}P_{n+1}(x) + \frac{n}{2n+1}P_{n-1}(x)$
for Legendre polynomials.

{\bf Note 2.} The Dirichlet-to-Neumann operator $\mathcal N$
(which correspond  the boundary condition $u_0$ to the normal
derivative of the field $u(r)$) could be constructed in the
following way:
$$
\left (\mathcal N u_0 \right ) (r) = \sum_{\bsn} \left (
\int_0^\pi \int_0^{2\pi} u_0(\theta,\varphi)
\overline{\widehat{\Psi}}_\bsn(r(\theta,\varphi))d\theta d \varphi
\right ) \frac{\partial \widehat{\Psi}_\bsn}{\partial n}(r).
$$
\subsection{Comparison with other explicit representations}
In spite of Theorem \ref{th} be stated for star shaped bodies
clearly all representations of the Green function will be the
same for arbitrary, smooth enough, body. We just should put
$$
 {u}_{\bsp}=\int_{0}^\pi \int_{0}^{2\pi} u_0(\theta,\varphi)
\overline{Y}_\bsp(\theta,\varphi) \overline{h}_{\bsp(1)}(k
|r(\theta,\varphi)|) d \theta \varphi, \quad \bsp \in \mathcal L,
$$
$$
 g_{\bsi \bsj} =\int_{0}^\pi \int_{0}^{2\pi}
Y_\bsi(\theta,\varphi)\overline{Y}_\bsj(\theta,\varphi)
h_{\bsi(1)}(k |r(\theta,\varphi)|)\overline{h}_{\bsj(1)}(k
|r(\theta,\varphi)|) d \theta \varphi, \quad \bsi,\bsj \in
\mathcal L.
$$
It is possible, like we mentioned above, functions
$\{Y_\bsp(\theta,\varphi) h_{\bsp(1)}(k |r(\theta,\varphi)|)\}$
do form a basis for arbitrary, smooth enough body.

Our representation holds for all values of $k>0$, but exists other
explicit representations for the Green's function which holds in a
neighborhoods of points $k=0$ and $k=\infty$. We consider of interest to cite the following
\begin{theorem}{R.~E.~ Kleinman \cite[Th. 4.1]{kleinm}}
There exists $\alpha>0$ such that when $|k|<\alpha$, the Green's
function $G(x,r)$ exists uniquely in $\Omega'$ and is given
explicitly by
\begin{equation}\label{nm}
G(r,t)=-\frac{e^{ik|r-t|}}{4\pi |r-t|} +e^{ikr} \sum_{n=0}^\infty
K^n U_0,
\end{equation}
where
$$
K(U_0)=-2ik \int_{\Omega'} dv(t_1) \frac{G_0(r,t_1)}{|t_1|}
\frac{\partial}{\partial |t_1|} [|t_1| U_0],
$$
$$
U_0=U_0(t_1,t)=\int_{\partial \Omega}
dS(t_0)\frac{e^{-ik|t_0|+ik|t_0-t|}}{4\pi |t_0-t|}
\frac{\partial}{\partial n_{t_0}}G_0(t_1,t_0).
$$
Here $v$ is a volume element, function $G_0(r,t_1)$ is the static
Dirichlet-Green function.
\end{theorem}
As we can see, all summands here are determined recursively as in
the Theorem \ref{th}. Also, is important to note, that the  first
approximation is a static Green function, which is supposed to be
known. So, in the some sense the constructed solution is a
pertubated static solution, that explain the quick, exponential,
convergence of the approximation (\ref{nm}).

In the high-frequency exists the following
representation for the Green's function which holds for obstacles
which satisfy non-trapping condition and single impact condition
(see \cite{AlberLeis} for details)
\begin{equation}\label{grH}
G(r,t)=-\frac{e^{ik|r-t|}}{4\pi |r-t|}+e^{ikS(r,t)} \sum_{m=0}^l
\left ( \frac i k \right ) z_m(r,t) + R_l(r,t),
\end{equation}
where function $S(r,t)$ satisfy {\it eikonal} equation
$$
| \nabla_r S|^2=1,
$$
 and
function $z_m(r,t)$ are defined recursively through differential
({\it transfer}) equations:
$$
2 \nabla_r S \cdot \nabla_r z_m +(\nabla_r S)z_m=-\nabla_r
z_{m-1}, \quad m \geq 0, z_{-1} \equiv 0.
$$
If  $r$ and $t$ don't lie on the tangential rays, we have
the estimate
$$
|R_l(r,t)| =o(k^{1-l}), \quad k \rightarrow \infty.
$$
In that case, the good, exponential,  convergence exists since
solution at high values of $k$ are close to the  ``classical'' limit
solution.

$\sigma_T$ representation (\ref{sigmat}) holds for all values of $k>0$, but we
don't have any estimates about the rate of convergence. For the
sphere, our representation is the well known solution (for ex.
\cite{morsef}) with very good convergence (as a $1/n!$). Of
course, good convergence properties should be kept in a some
neighborhood of the sphere. Moreover, we are sure that it will be
possible to do it with developing of local limit theorems for {\it sign}
distributions, really if $1/r$ is a characteristic function of a
random variable with a sign distribution then elements of a Gram
matrix $g_{\bsi \bsj}$ are in the sense of the probability of sum of
$(\bsi +\bsj +2)$ of such variables (see \cite{hoch} for integral
theorem). By now it is out of our possibilities  to
describe the neighborhood of at least exponential convergence.
From another point of view we want to remind that representation
(\ref{sigmat}), in contrast to representations
(\ref{grH}, \ref{nm}), converges for all values of $k>0$.

To finish our discussion we want to note that representation like
(\ref{sigmat}) can not be done for the 2dimensional model, since
Bessel functions of integer index do not allow a finite polynomial
representation.

\subsection{Polyhedrons and bodies which are revolutions of polylines}

Calculation of the Gram matrix $G=\{g_{ij}\}$ for
polyhedron is very natural and elements $g_{ij}$ need only $N\,
(i+j)^2$ sums, where $N$ is the number of sides.

Let's set $N+1$ points on the segment $[0,\pi]$
$$
0=\theta_0 \leq \theta_1  \leq \ldots \leq \theta_{N-1} \leq \theta_N = \pi,
$$
and consider the set of bodies $\mathcal R$ which contains arbitrary bodies of
revolution of the polyline with vertexes in points $\{\theta_i,
i=0,\ldots,N\}$.

\begin{theorem}\label{poly}
Consider polyline $r=\sum_i r_i(\theta)I_i(\theta)$, where $I_i, i=1,\ldots, N$
is a characteristic function of the segment $[\theta_{i-1},\theta_i]$ and
coefficients $\{a_i,b_i,c_i,  i=1,\ldots,N \}$ determine parts of the polyline
$r_i$:
\begin{equation}\label{defpol}
a_i r_i \cos \theta +b_i r_i \sin \theta= f_i, \quad  i=1,\ldots, N.
\end{equation}
Then Gram matrix coefficients $g_{i j}$ could be calculated as\\
$$
g_{i j}=\sum_{p=1}^N  \frac{1}{(f_p)^{i+j+2}}\sum_{l=0}^{i+j+2} C_{i+j+2}^l
a_p^l b_p^{i+j+2-l} T^p_{i j,l (i+j)},
$$
where
$$
T^p_{i j,l m}=\int_{\theta_{p-1}}^{\theta_p} Y_{i}(\theta)
\overline{Y}_j(\theta) \cos^l \theta \sin^{m+2-l} \theta  \sin \theta d\theta,
\quad l \leq m+2;
$$
$$
g_{i j}=\sum_{p=1}^N \sum_{m=0}^{i+j}\frac{1}{k^{m+2}} \left (\sum_{n=0}^{m}
\widehat{h}_{i n} \overline{\widehat{h}}_{j (m-n)} \frac{1}{(f_p)^{m+2}}\right
) \left (  \sum_{l=0}^{m+2} C_{m+2}^l a_p^l b_p^{m+2-l} T^p_{i j,l m} \right )\,.
$$
\end{theorem}

\subsection{Explicit dependence of Gram matrix on frequency}

 We can represent elements of the Gram matrix $\{g_{\bsi \bsj},
\bsi,\bsj \in \mathcal L \}$ for arbitrary smooth enough surface
 as a polynomial of the inverse frequency $k^{-1}$,
\begin{equation}\label{lc1}
g_{\bsi \bsj}= (-1)^{j(2)}\sum_{m=0}^{i(1)+j(1)} \frac{1}{k^{m+2}} p_{\bsi
\bsj}^m,
\end{equation}
where
\begin{equation}\label{lc2}
 p_{\bsi
\bsj}^m= \sum_{l=0}^{m} \widehat{h}_{\bsi(1)l}
\overline{\widehat{h}}_{\bsj(1)(m-l)}  \int_{0}^{\pi} d \theta
\int_{0}^{2\pi} d\varphi \frac{Y_\bsi(\theta,\varphi)
{Y}_{\overline{\bsj}}(\theta,\varphi)}{|r(\theta,\varphi)|^{m+2}},
\quad 0 \leq m \leq i(1)+j(1).
\end{equation}


\section{Proofs}

Denote
$$\Psi_\bsn(r)=Y_\bsn(\theta)h_{\bsn(1)}(k|r|), \quad \bsn \in \mathcal L, \quad r \in \Omega',$$
where functions $\Psi_\bsn(r)$ satisfy (\ref{helm}) and
(\ref{Somm}). Moreover, they have asymptotics at the infinity (see
(\ref{hankdef}))
$$
\Psi_\bsn(r) \sim \frac{1}{k}Y_\bsn(\theta) \frac{e^{ik|r|}}{|r|},
\quad |r| \rightarrow \infty\,.
$$
Let $A$ be a linear operator acting from $L_2(\partial \Omega,dS)$
to $L_2(S^2, d\sigma)$  which corresponds $u_{|\partial \Omega}$
to scattering amplitude $f(q)$ (see (\ref{scamp})). Here $dS$ and
$d \sigma$ are standard metrics on $\partial \Omega$ and unit
sphere $S^2$. In \cite[2.12,2.16]{AlRammGr},
\cite[2.10,2.13]{AlRammPhysA} it is proved that $A$ is a bounded
operator and in particulary it is proved that functions
${\Psi_{\bsn}}_{|\partial \Omega}$ form  a basis in $L_2(\partial
\Omega,dS)$. So we have the transformation
$$
A \left ( \sum_{\bsn \in \mathcal L} c_\bsn \Psi_\bsn \right) = \frac{1} {k}
\sum_{\bsn \in \mathcal L} c_\bsn Y_\bsn, \quad \sum_{\bsn \in \mathcal L}
c_\bsn \Psi_\bsn \in L_2(\partial \Omega,dS).
$$
Let us prove that exists a constant $C_1=C_1(k,\Omega)$ such that
\begin{equation}\label{est1}
|c_\bsn| \leq \frac{C_1}{\bsn(1)!}, \quad \bsn \in \mathcal L.
\end{equation}
Consider sphere $S_R:=\{|r|=R\}$ such that the body $\Omega$ is
strictly embedded in that sphere. Define also the $A_1 :
L_2(\partial \Omega,dS) \rightarrow L_2(S_R, d \sigma_R)$, where
$d \sigma_R =R^2 \sin \theta d \theta d \varphi$, as
$$
(A_1 u_0)(x)=\int_{\partial \Omega} \frac{\partial G(x,y)}{\partial n} u_0(y)
dS(y), \quad x \in S_R.
$$
Since function $\frac{\partial G}{\partial n}$ has  singularities
only for $x=y$ (for example, see \cite{kleinm}) operator $A_1$ is
bounded. Note that functions $h_\bsn(k|r|)$ are constant on the
$S_R$,
$$
4\pi R^2 |h_\bsn(kR)| \cdot |c_{\bsn}|=|(A_1 u_0,\overline{Y}_\bsn)_{L_2(S_R)}|
\leq
$$
$$
 \|A_1 u_0\|_{L_2(S_R)}
\|\overline{Y}_\bsn\|_{L_2(S_R)}=4 \pi R^2 \|A_1 u_0\|_{L_2(S_R)}
\leq 4 \pi R^2 \|A_1 \| \|u_0\|_{L_2(\partial \Omega)}.
$$
So
$$
|c_{\bsn}| \leq \frac{\|A_1 \| \|u_0\|_{L_2(\partial
\Omega)}}{|h_\bsn(kR)|}
$$
taking into account the asymptotic (see \cite[pp. 358-364]{abram})
of $h_n(R) \sim \frac{n!}{(R/2)^n}$, $n \rightarrow \infty$, and
since $R$ could be chosen to satisfy $kR \geq 2$, we obtain
(\ref{est1}) (that corresponds to (\ref{estth1} in Theorem 1)).

Let $g(\theta,\varphi)$ be a density
\begin{equation}\label{predg}
dS=[g(\theta,\varphi)]^2  d \sigma = |r|\sqrt{ \left [ |r|^2+\left
( \frac{\partial |r|}{\partial \theta} \right )^2 \right ] \sin^2
\theta + \left ( \frac{\partial |r|}{\partial \varphi} \right )^2}
d\theta d\varphi
\end{equation}
and for simplicity we shall use the same notation $d \sigma= d
\theta d \varphi$ for measure on the $\partial \Omega$. It is
evident that if function $f \in L_2(\partial \Omega,d \sigma)$
then $f/g \in L_2(\partial \Omega,d S)$.

 Denote $\Psi_\bsn^0=\Psi_\bsn |_{\partial \Omega} / g$ and
construct an orthonormal basis $\wpsi^0_n$ in $L_2(\partial
\Omega,dS)$.  We shall construct it in form:
\begin{equation}\label{pred}
\wpsi^0_\bsn=\sum_{\bsk=\bso}^\bsn c_{\bsn \bsk} \Psi_\bsk^0,
\quad \bsn \in \mathcal L
\end{equation}
$$
\wpsi^0_\bsn=[\Psi_\bsn^0-\sum_{\bsk=\bso}^{\bsn-1}
(\Psi_\bsn^0,\wpsi^0_\bsk)\wpsi^0_\bsk]/\lam_\bsn
$$
where
$$
\lam_\bsn={\left[ \|\Psi_\bsn^0\|^2 -\sum_{\bsk=\bso}^{\bsn-1}
|(\Psi_\bsn^0,\wpsi^0_\bsk)|^2 \right]^{1/2}}\,.
$$
Using
$$
(\Psi_\bsn^0,\wpsi^0_\bsk)=\sum_{\bsp=\bso}^\bsk
\overline{c}_{\bsk \bsp} (\Psi_\bsn^0,\wpsi^0_\bsp)
$$
we have
$$
c_{\bsn \bsn}=1/\lam_\bsn, \quad c_{\bsn \bsm}=\left
[\sum_{\bsk=\bsm}^{\bsn-1} \sum_{\bsp=\bso}^{\bsk}
\overline{c}_{\bsk \bsp} c_{\bsk \bsm} (\Psi_\bsn^0,\Psi_\bsp^0)
\right ]/\lam_\bsn, \quad \bsm<\bsn\,.
$$
Denote
$$
\wpsi_\bsn(r)=g\wpsi^0_\bsn=\sum_{\bsk=\bso}^\bsn c_{\bsn \bsk}
\Psi_\bsk, \quad \bsn \in \mathcal L, \quad r \in \Omega '.
$$
and let $u_0$ be a function in $L_2(\partial \Omega,d\sigma)$ and
also $u_0/g=\sum_0^\infty \widehat{u}_\bsn \wpsi^0_\bsn$. Then
function
$$
u(r)=\sum_{\bsn=\bso}^\infty \widehat{u}_\bsn \wpsi_\bsn(r), \quad
r \in \mathbb R^3 \backslash \Omega
$$
satisfy (\ref{helm})-(\ref{dir}).

Using (\ref{pred}) we obtain the corresponding scattering
frequencies (see (\ref{defAm})),
$$
A_\bsm=\frac{1}{k} \sum_{\bsn=\bsm}^\infty  c_{\bsn \bsm} \left (
\sum_{\bsp=\bso}^\bsn \overline{c}_{\bsn \bsp}  u_\bsp \right ),
$$
where $u_\bsp=(u_0/g,\Psi_\bsp^0)_{L_2(\partial \Omega,
dS)}=(u_0,\left . \Psi_\bsp \right |_{\partial
\Omega})_{L_2(\partial \Omega, d\sigma)}$.

By construction $\sum_{\bsp=\bso}^\bsn \overline{c}_{\bsn \bsp}
u_\bsp = \widehat{u}_\bsn$ and that follows (\ref{estth2}).

Let us prove now (\ref{estth3}). Set
$$
\lam_\bsn= \|F_\bsn\|_{L_2(\partial \Omega, dS)}, \quad \mbox{
where } F_\bsn=\Psi_\bsn^0-\sum_{\bsk=\bso}^{\bsn-1}
(\Psi_\bsn^0,\wpsi^0_\bsk)\wpsi_\bsk^0\,.
$$
By construction $(A F_\bsn, Y_\bsn) = \frac{1}{k}$, therefore $\|
A F_\bsn \| \geq \frac{1}{k}$, so
$$
\frac{1}{k} \leq \| A F_\bsn \| \leq \|A\|
\|F_\bsn\|_{L_2(\partial \Omega, dS)}\,.
$$
Due to the representation of $r(\theta,\varphi)$ and (\ref{predg})
exists $S=\max_{\theta,\varphi} |g(\theta,\varphi)|^2<\infty$, so
we have
$$
\lam_\bsn =\|F_\bsn\|_{L_2(\partial \Omega, d\sigma )} \geq
\frac{1}{S} \|F_\bsn\|_{L_2(\partial \Omega, d S )} \geq (S \|A\|
k)^{-1}\,.
$$

Now, let us prove (\ref{sigmat}). 
We write (see (\ref{amam}))
\begin{eqnarray*}
|A_\bsm|^2 & = & \frac{1}{k^2} \sum_{\bsn \geq \bsm} \left [c_{\bsn
\bsm} \sum_{\bsp \leq \bsn} \overline{c}_{\bsn \bsp} \widehat{u}_p
\left (c_{\bsn \bsm}  \sum_{\bsp \leq \bsn} \overline{c}_{\bsn
\bsp} \widehat{u}_p+2\sum_{\bsm <\bsl < \bsn} c_{\bsl \bsm}
\sum_{\bsp \leq \bsl} \overline{c}_{\bsl \bsp} \widehat{u}_p\right
) \right ],\\
\sigma_T & = & \frac{1}{k^2} \sum_{\bsn=\bso}^\infty\sum_{\bsm
\leq \bsn} \left [\overline{c}_{\bsn \bsm} \left ( \sum_{\bsp \leq
\bsn} c_{\bsn \bsp} \overline{\widehat{u}}_p \right )\left
(c_{\bsn \bsm} \sum_{\bsp \leq \bsn} \overline{c}_{\bsn \bsp}
\widehat{u}_p+2\sum_{\bsm <\bsl < \bsn} c_{\bsl \bsm} \sum_{\bsp
\leq \bsl} \overline{c}_{\bsl \bsp} \widehat{u}_p\right ) \right
].
\end{eqnarray*}
Next, we calculate $(\Psi_\bsi^0,\Psi^0_\bsj)$,
$$
g_{\bsi \bsj}= (\Psi_\bsi^0,\Psi^0_\bsj)=\int_{0}^{\pi} d \theta
\int_{0}^{2\pi} d\varphi Y_\bsi(\theta,\varphi)
\overline{Y}_\bsj(\theta,\varphi)
h_{\bsi(1)}(k|r(\theta,\varphi)|)
\overline{h}_{\bsj(1)}(k|r(\theta,\varphi)|) =
$$
$$
(-1)^{j(2)} \int_{0}^{\pi} d \theta \int_{0}^{2\pi} d\varphi
Y_\bsi(\theta,\varphi) Y_{\overline{\bsj}}(\theta,\varphi) \left (
\sum_{p=0}^{\bsi(1)}\frac{\widehat{h}_{\bsi(1)p}}{(k|r|)^{p+1}}
\right ) \left (
\sum_{p=0}^{\bsj(1)}\frac{\overline{\widehat{h}}_{\bsj(1)p}}{(k|r|)^{p+1}}
\right )=
$$
$$
(-1)^{j(2)}  \int_{0}^{\pi} d \theta \int_{0}^{2\pi} d\varphi
Y_\bsi(\theta,\varphi) {Y}_{\overline{\bsj}}(\theta,\varphi)
\sum_{m=0}^{i(1)+j(1)}\frac{1}{k^{m+2}}\left (\sum_{l=0}^{m}
\widehat{h}_{\bsi(1)l} \overline{\widehat{h}}_{\bsj(1)(m-l)}
\right ) \frac{1}{|r(\theta,\varphi)|^{m+2}}=
$$
\begin{equation}\label{lc0}
(-1)^{j(2)}\sum_{m=0}^{i(1)+j(1)}\frac{1}{k^{m+2}} \sum_{l=0}^{m}
\widehat{h}_{\bsi(1)l} \overline{\widehat{h}}_{\bsj(1)(m-l)}
\int_{0}^{\pi} d \theta \int_{0}^{2\pi} d\varphi
\frac{Y_\bsi(\theta,\varphi)
{Y}_{\overline{\bsj}}(\theta,\varphi)}{|r(\theta,\varphi)|^{m+2}}=
\end{equation}
$$
(-1)^{\bsj(2)}\sum_{m=0}^{\bsi(1)+\bsj(1)} \frac{1}{k^{m+2}}\left
(\sum_{l=0}^{m} \widehat{h}_{\bsi(1)l}
\overline{\widehat{h}}_{\bsj(1)(m-l)} \right ) \sum_{d: |d| =m+2,
\supp{d} \leq
 (N,N)} C^d a^d I^{d+e_\bsi+e_{\overline \bsj }}.
$$
Theorem \ref{th} is proved.

\textbf{Proof for Polyhedrons --- Coefficients $g_{i j}$}.
From
(\ref{defpol}) we have
$$
\frac{1}{r_i} =\frac{a_i \cos \theta +b_i  \sin \theta}{f_i}, \quad i=1,\ldots,
N.
$$
First of all note that
$$
g_{i j}=\sum_{p=1}^N g_{i j}^p,\textrm{ with } g_{i j}^p :=
\int_{x_{p-1}}^{x_p} \Psi_{i}(r_p(\theta),\theta)
\overline{\Psi}_{j}(r_p(\theta),\theta) \sin \theta d\theta.
$$
Develop $g_{i j}^p$,
$$
g_{i j}^p= \int_{x_{p-1}}^{x_p} Y_{i}(\theta)
\overline{Y}_j(\theta) h_i(kr(\theta))h_j(k|r|(\theta))\sin \theta
d\theta =
$$
$$= \sum_{m=0}^{i+j}\frac{1}{k^{m+2}} \sum_{l=0}^{m}
\widehat{h}_{i(1) l} \overline{\widehat{h}}_{j (m-l)}
\int_{0}^{\pi} d \theta \frac{Y_i(\theta,\varphi)
{Y}_{j}(\theta,\varphi)}{|r(\theta,\varphi)|^{m+2}}=
$$
$$= \sum_{m=0}^{i+j}\frac{1}{k^{m+2}} \sum_{l=0}^{m}
\widehat{h}_{i(1) l} \overline{\widehat{h}}_{j (m-l)}
\frac{1}{(f_p)^{m+2}}\sum_{l=0}^{m+2} C_{m+2}^l a_p^l b_p^{m+2-l} T^p_{i j,l
m},
$$
where
$$
T^p_{i j,l m}=\int_{x_{p-1}}^{x_p} Y_{i}(\theta)
\overline{Y}_j(\theta) \cos^l \theta \sin^{m+2-l} \theta  \sin
\theta d\theta, \quad l \leq m+2.
$$

Proof of the representation (\ref{lc1}) follows explicitly from
(\ref{lc0}).

\vspace{8mm}

\textbf{Acknowledgments:} E.~L.~Lakshtanov is thankful to A.~G.~Ramm for many useful discussions.
This work was supported by {\it Centre for Research on Optimization and
Control} (CEOC) from the ``{\it Funda\c{c}\~{a}o para a
Ci\^{e}ncia e a Tecnologia}'' (FCT), cofinanced by the European
Community Fund FEDER/POCTI, and by the FCT research project
PTDC/MAT/72840/2006.

\end{document}